\documentclass[sigconf]{acmart}

\usepackage{booktabs} 
\usepackage{amsmath}
\usepackage{listings}
\usepackage{xspace}
\usepackage{caption}
\usepackage{subcaption}
\usepackage{balance}

\graphicspath{{figure/}}

\copyrightyear{2019}
\acmYear{2019} 
\setcopyright{iw3c2w3}
\acmConference[WWW '19]{Proceedings of the 2019 World Wide Web Conference}{May 13--17, 2019}{San Francisco, CA, USA}
\acmBooktitle{Proceedings of the 2019 World Wide Web Conference (WWW '19), May 13--17, 2019, San Francisco, CA, USA}
\acmPrice{}
\acmDOI{10.1145/3308558.3314139}
\acmISBN{978-1-4503-6674-8/19/05}

\begin{document}
\title[CityFlow]{CityFlow: A Multi-Agent Reinforcement Learning Environment for Large Scale City Traffic Scenario}

\author{Huichu Zhang}
\email{zhc@apex.sjtu.edu.cn}
\affiliation{%
  \institution{Shanghai Jiao Tong University}
  \streetaddress{800 Dongchuan Road}
  \city{Shanghai}
  \country{China}
}

\author{Siyuan Feng}
\email{hzfengsy@sjtu.edu.cn}
\affiliation{%
  \institution{Shanghai Jiao Tong University}
  \streetaddress{800 Dongchuan Road}
  \city{Shanghai}
  \country{China}
}

\author{Chang Liu}
\email{only-changer@sjtu.edu.cn}
\affiliation{%
  \institution{Shanghai Jiao Tong University}
  \streetaddress{800 Dongchuan Road}
  \city{Shanghai}
  \country{China}
}
\author{Yaoyao Ding}
\email{yyding@sjtu.edu.cn}
\affiliation{%
  \institution{Shanghai Jiao Tong University}
}
\author{Yichen Zhu}
\email{zyc_IEEE@sjtu.edu.cn}
\affiliation{%
  \institution{Shanghai Jiao Tong University}
}
\author{Zihan Zhou}
\email{footoredo@sjtu.edu.cn}
\affiliation{%
  \institution{Shanghai Jiao Tong University}
}
\author{Weinan Zhang}
\authornote{Corresponding author}
\email{wnzhang@sjtu.edu.cn}
\affiliation{%
  \institution{Shanghai Jiao Tong University}
}

\author{Yong Yu}
\email{yyu@apex.sjtu.edu.cn}
\affiliation{%
  \institution{Shanghai Jiao Tong University}
}

\author{Haiming Jin}
\email{jinhaiming@sjtu.edu.cn}
\affiliation{%
  \institution{Shanghai Jiao Tong University}
}

\author{Zhenhui Li}
\authornote{Corresponding author}
\email{jessieli@ist.psu.edu}
\affiliation{%
  \institution{Pennsylvania State University}
  \streetaddress{Old Main}
  \city{State College}
  \state{Pennsylvania}
  \country{USA}
}

\renewcommand{\shortauthors}{H. Zhang et al.}

\newcommand{\name}{\textsf{CityFlow}\xspace}
\newcommand{\nop}[1]{}

\begin{abstract}
Traffic signal control is an emerging application scenario for reinforcement learning. Besides being as an important problem that affects people's daily life in commuting, traffic signal control poses its unique challenges for reinforcement learning in terms of adapting to dynamic traffic environment and coordinating thousands of agents including vehicles and pedestrians. A key factor in the success of modern reinforcement learning relies on a good simulator to generate a large number of data samples for learning. The most commonly used open-source traffic simulator SUMO is, however, not scalable to large road network and large traffic flow, which hinders the study of reinforcement learning on traffic scenarios. This motivates us to create a new traffic simulator \name with fundamentally optimized  data structures and efficient algorithms. \name can support flexible definitions for road network and traffic flow based on synthetic and real-world data. It also provides user-friendly interface for reinforcement learning. Most importantly, \name is more than twenty times faster than SUMO and is capable of supporting city-wide traffic simulation with an interactive render for monitoring. Besides traffic signal control, \name could serve as the base for other transportation studies and can create new possibilities to test machine learning methods in the intelligent transportation domain.
\end{abstract}



\maketitle


\section{introduction}
Traffic signal control problem, one of the biggest urban problems, is drawing increasing attention in recent years~\cite{wei2018intellilight, liLW16, van2016coordinated}. 
Recent advances are enabled by large-scale real-time traffic data collected from various sources such as vehicle tracking device, location-based mobile services, and road surveillance cameras through advanced sensing technology and web infrastructure.  Traffic signal control is interesting but complex because of the dynamics of traffic flow and the difficulties to coordinate thousands of traffic signals. Reinforcement learning becomes one of the promising approaches to optimize traffic signal plans, as shown in several recent studies~\cite{wei2018intellilight, liLW16, van2016coordinated}. At the same time, traffic signal control is also one of the major real-world application scenarios for reinforcement learning~\cite{li2017deep}. 

To successfully deploy reinforcement learning technique for traffic signal control, the traffic simulator becomes the most important factor. Because the learning method relies on a large set of data samples. These data samples can hardly be collected from the real world directly. Aside from the consequence of bad decisions, a city simply cannot generate enough data samples for learning. If we treat each minute as a data sample, a city can only generate 1,440 (24 hours by 60 minutes) data samples in a day. Such a small size of sample is not enough to train a deep reinforcement learning model to be powerful enough to make good decisions. Thus, it becomes crucial to have a simulator that is fast enough to generate a large set of data samples. 

The most popular public traffic simulator SUMO~\cite{SUMO2018} (Simulation of Urban Mobility) has been frequently used in many recent studies. SUMO, however, is not scalable to the size of the road network and the size of traffic flow. For example, it can only perform around three simulation steps per second on a $30\times 30$ grid with tens of thousands of vehicles, the situation is even worse if we use the python interface to get information about the system to support reinforcement learning. A city, however, is often at the size of a thousand intersections (e.g. there are $30\times 40$ intersections of major roads in Hangzhou, China) and hundreds of thousands vehicles, which is beyond the current simulation capacity of SUMO. 

To enable the reinforcement learning for intelligent transportation, we create a traffic simulator \name\footnote{https://github.com/cityflow-project/CityFlow/}, which can be scaled to support the city-wide traffic simulation. One of the major improvements over SUMO is that \name enables multithreading computing.  To the best of our knowledge, this is the first open-source simulator that can support city-wide traffic simulator. \name is flexible to define road network, vehicle models, and traffic signal plans.  It is more than twenty times faster than SUMO. We have also provided friendly interface for reinforcement learning testbed. We plan to demonstrate these functions at the demo session.

Finally, our scalable traffic simulator \name will open many new possibilities besides traffic signal control scenario. First, it could support various large-scale transportation research studies, such as vehicle routing through mobile app, traffic jam prevention. Second, similar to OpenAI Gym\footnote{\url{https://gym.openai.com/}} which provides a set of benchmark environments for reinforcement learning, \name could serve as a benchmark reinforcement learning environment for transportation studies. Besides traffic signal control, reinforcement learning has been used in transportation studies such as taxi dispatching~\cite{xu2018large} and mixed autonomy systems~\cite{wu2017emergent}. But all the existing studies either use SUMO or over-simplified traffic simulator. Third, we plan to better calibrate the simulation parameters by learning from real-world observations. This will make the simulator not only generate data samples fast but also generate ``real'' data samples.

\section{Brief Description}

\subsection{System Design}
\name is a microscopic traffic simulator which simulates the behavior of each vehicle at each time step, providing highest level of detail in the evolution of traffic. However, microscopic traffic simulators are subject to slow simulation speed \cite{yin2011comparison}. Unlike SUMO, \name uses multithreading to accelerate the simulation. Data structure and simulation algorithm are also optimized to further speedup of the process.

\subsubsection{Road Network}
Road network is the basic data structure in \name. \textbf{Road} represents a directional road from one \textbf{intersection} to another \textbf{intersection} with road-specific properties. A \textbf{road} may contain multiple \textbf{lanes}. Each \textbf{lane} holds a Linked List of vehicles. Linked List supports fast insertion and searching of leading vehicles. \textbf{Segments} are small fragments of a \textbf{lane}. We design segments in order to efficiently find all vehicles within a certain range of the lane. This structure is crucial for fast lane change operation. \textbf{Intersection} is where roads intersects. An \textbf{intersection} contains several \textbf{roadlinks}. Each roadlink connects two roads of the intersection and can be controlled by traffic signals. A \textbf{roadlink} contains several \textbf{lanelinks}. Each lanelink represents a specific path from one lane of incoming road to one lane of outgoing road. \textbf{Cross} represents the cross point between two lanelinks. This structure is crucial for fast intersection logic.

\subsubsection{Car Following Model}
The car-following model is the core component \name. It computes the desired speed of each vehicle at next step using information like traffic signal, leading vehicles, etc. and ensures that no collisions occur in the system. Currently, the car following model used in \name is a modification of the model proposed by Stephen Krau\ss~\cite{krauss1998microscopic}. The key idea is that: the vehicle will drive as fast as possible subject to perfect safety regularization (e.g. being able to stop even if leading vehicle stops using maximum deceleration). Unlike SUMO~\cite{SUMO2018}, we use ballistic position update rule instead of Euler position update. Ballistic update yields more realistic dynamics for car-following models based on continuous dynamics especially for larger time-steps (e.g. 1 second) \cite{treiber2015comparing}.

Basically, vehicles are subject to several speed constraints, maximum speed which meets all these constraints will be chosen. Currently, following constraints are considered:
\begin{itemize}
\item vehicle and driver's maximum acceleration
\item road speed limit
\item collision free following speed
\item headway time following speed
\item intersection related speed
\end{itemize}

Due to page limit, we only present the detail of collision free speed computation. It takes $v_F$ current speed of following vehicle, $v_L$ current speed of leading vehicle, $d_F$ maximum deceleration of following vehicle, $d_L$ maximum deceleration of leading vehicle, $gap$ current gap between two vehicles, $interval$ the length of each time step as parameters and compute the no-collision-speed $s$ by solving a quadratic equation using equation \ref{eqn:speed}.
\begin{align}
c & = \frac{v_F\cdot interval}{2} - \frac{v_L^2}{2\cdot d_L} - gap \notag \\
a & = \frac{1}{2\cdot d_F} \notag \\
b & = \frac{interval}{2} \notag \\
s & = \frac{-b+\sqrt{b^2-4\cdot a\cdot c}}{2\cdot a} \label{eqn:speed}
\end{align}

Intersection related speed is handled by intersection logic and is illustrated in the next section.

\subsubsection{Intersection Logic}
The behavior of vehicles in intersection is complex and it requires careful design to efficiently mimic real world behavior \cite{krajzewicz2013road,fellendorf2010microscopic}. Basically, vehicles in intersection should obey following two rules:
\begin{itemize}
\item fully stop at red signal, stop if possible at yellow signal
\item yield to vehicles with higher priority (e.g. turning vehicles should yield straight-moving vehicles)
\end{itemize}
To avoid collisions at intersection, it is non-trivial to check if there are vehicles on the opposite lane. The simplest method is to use brute force search to find all vehicles within a certain range and check if they will collide within a certain time period. But this method is very time consuming. Instead, we precompute all the cross points between lanelinks in intersection. When a vehicle approaches the intersection, it will notify all cross points in the intersection about its arrival. The cross points is responsible for deciding which vehicle could pass and which vehicle should yield. The time complexity of our algorithm is $\mathcal{O}(N_{crosspoints})$. Due to page limit, we omit the detail of our algorithm.

\subsubsection{Lane Change Model}
Lane change model addresses two questions for a vehicle: when and how to change lane.
Vehicles may change lanes when there are more free space on adjacent lanes or a lane change is required to follow its route. Notice that it is slow to traverse all vehicles in adjacent lanes. Instead, by maintaining the vehicle information in \textbf{segments} which are small fragments of each lane, we only need to search for related vehicles in adjacent segments in constant time (up to three segments for each lane), which largely reduce time complexity.

When a vehicle decides to change lane, it needs to find a way to notify other vehicles. Here we use a similar mechanism in SUMO. When a vehicle changes lane, the simulation engine will put a copy of it to its destination lane, called shadow vehicle. A shadow vehicle has the same function as a normal vehicle, and it can become the leader of other vehicles in the car following model. The vehicle and its shadow moves consistently, which is guaranteed by the simulation engine in the way that their speed constraints will be applied to each other. After the lane change finishes, the simulation engine will just remove the original vehicle and let its shadow vehicle replace it.

\subsection{Python Interface}
In order to support multi-agent reinforcement learning, we provide a python interface via \textit{pybind11}~\cite{pybind11}. User can perform simulation step by step and get various kinds of information about current state, e.g. number of vehicles on lane, speed of vehicles. Besides, we provide interface to control the elements in simulator at each time step. Currently, users can control traffic signals and add vehicles on-the-fly. We plan to support more types of controlling functions such as vehicle behavior control and road property control in the future. Below is a sample usage of python interface.

\begin{lstlisting}[language=Python, caption=usage of python interface, frame=tlrb, breaklines=true]
import engine
eng = engine.Engine(config_file)
phase = [...] # the traffic signal phase of each time step
for step in range(3600):
    eng.set_tl_phase("intersection_1_1", phase[step])
    eng.next_step()
    eng.get_current_time()
    eng.get_lane_vehicle_count()
    eng.get_lane_waiting_vehicle_count()
    eng.get_lane_vehicles()
    eng.get_vehicle_speed()
    # do something
\end{lstlisting}

\subsection{Frontend}
\begin{figure*}
    \centering
    \begin{subfigure}[b]{0.6\textwidth}
        \centering
        \includegraphics[width=\textwidth]{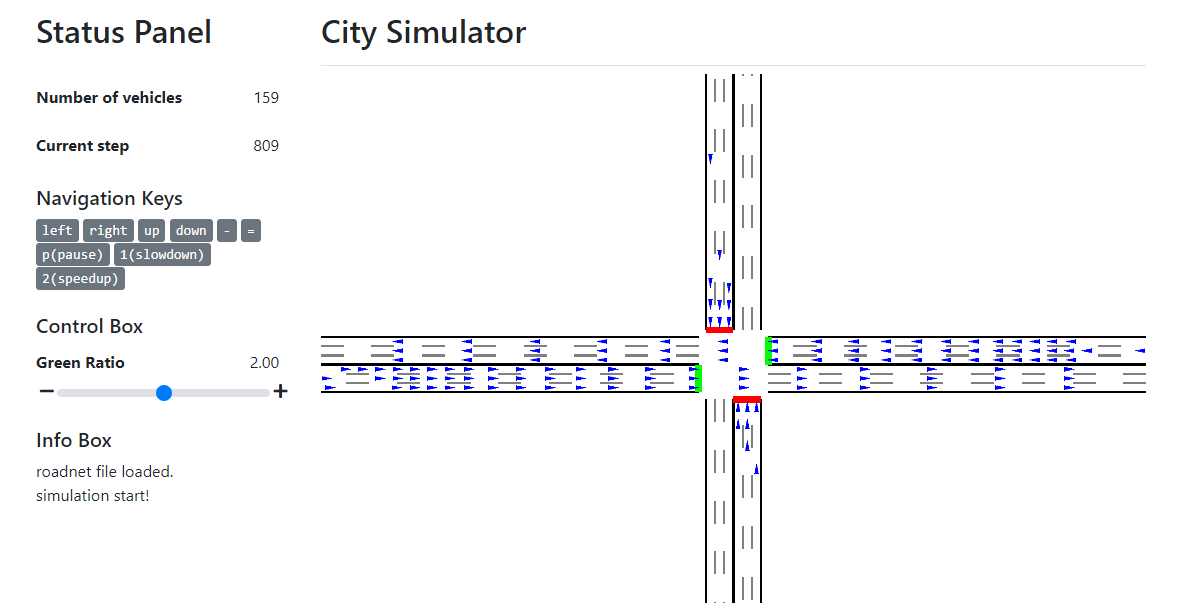}
        \caption{{\small $1\times 1$ road network with different green ratio}}
    \end{subfigure}
    \quad
    \begin{subfigure}[b]{0.3\textwidth}  
        \centering 
        \includegraphics[width=0.85\textwidth]{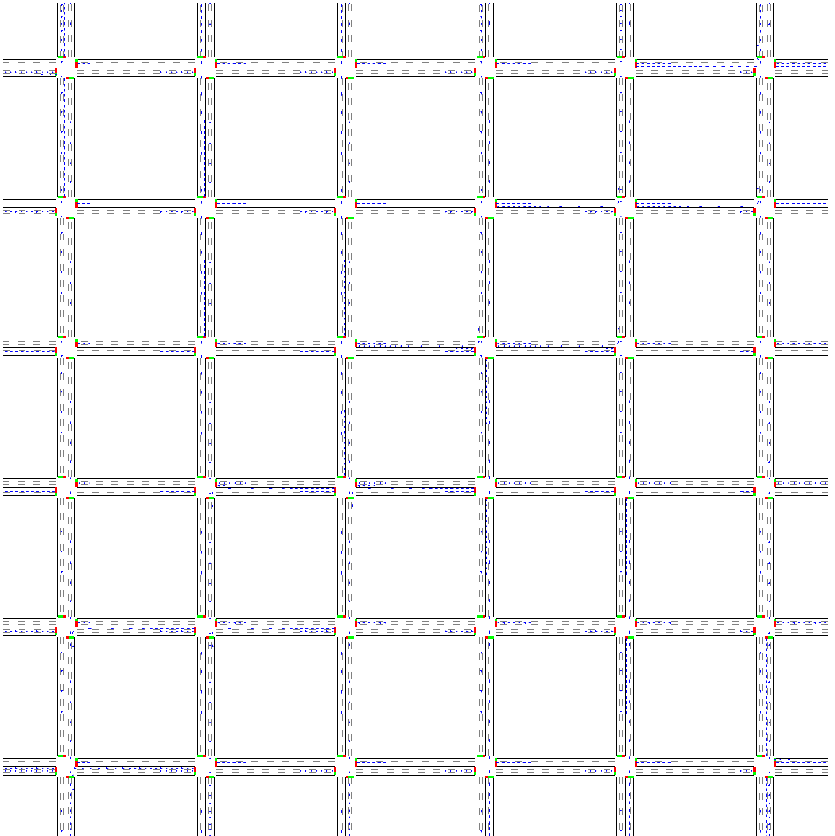}
        \caption{{\small $6\times 6$ grid road network}}
    \end{subfigure}
    \vskip\baselineskip
    \begin{subfigure}[b]{0.6\textwidth}   
        \centering 
        \includegraphics[width=\textwidth]{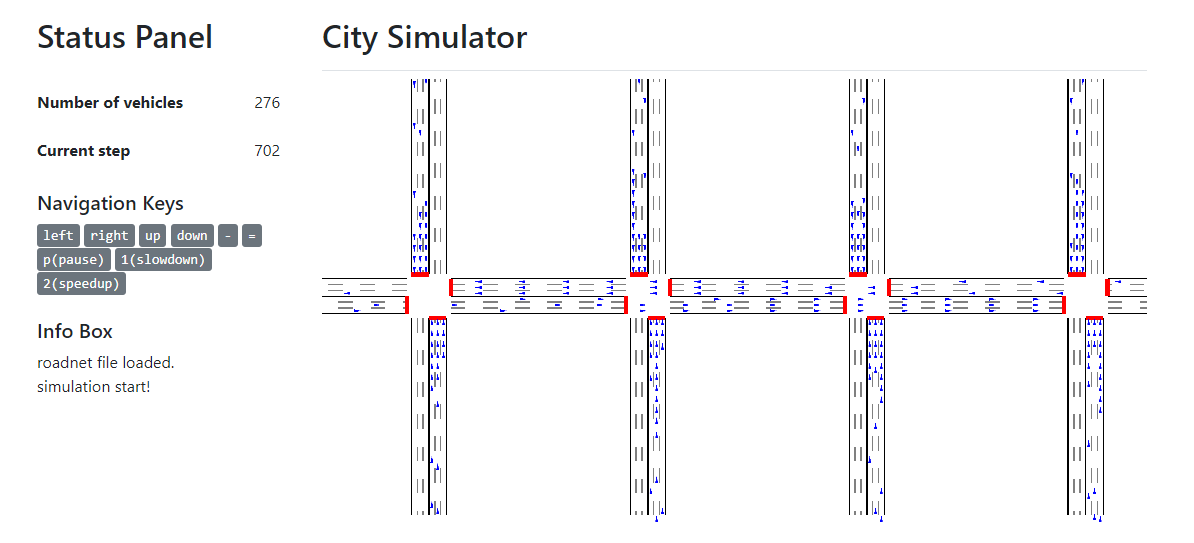}
        \caption{{\small $1\times 4$ grid road network}}
    \end{subfigure}
    \quad
    \begin{subfigure}[b]{0.3\textwidth}   
        \centering
        \includegraphics[width=0.85\textwidth]{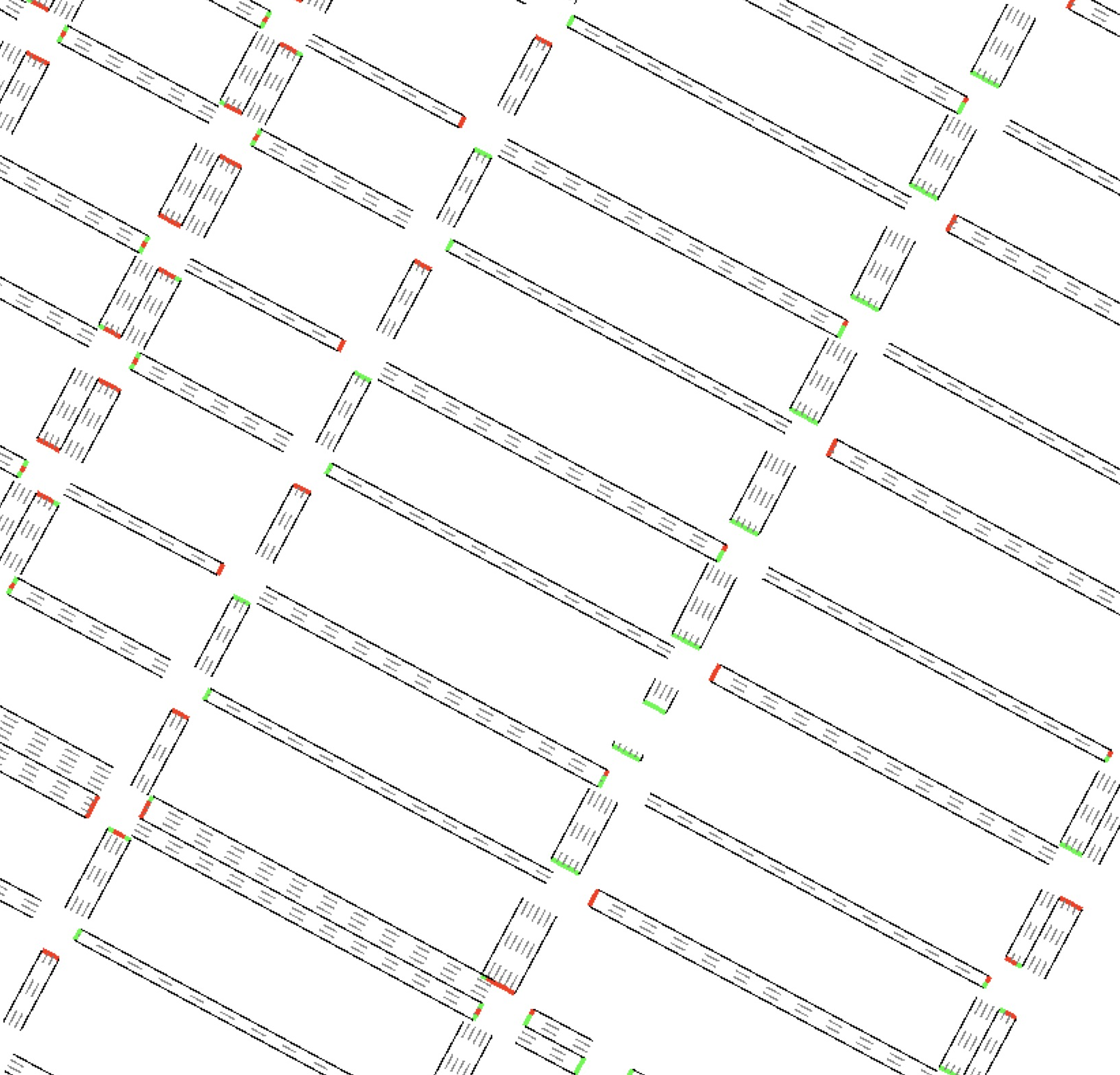}
        \caption{{\small Part of Manhattan road network}}
    \end{subfigure}
    \caption{Screenshot of \name in different scenario}
    \label{fig:screenshot}
\end{figure*}

We provide a web-based Graphic User Interface. User can check the replay output by the simulator. In order to support viewing large-scale simulation, we use \textit{WebGL}-based library \textit{PixiJS}\footnote{https://github.com/pixijs/pixi.js} for fast rendering of vehicles and traffic signals. Figure \ref{fig:screenshot} shows some screenshots of the GUI under several scenarios.

\section{Performance}
\begin{figure}[htb]
  \includegraphics[width=\columnwidth]{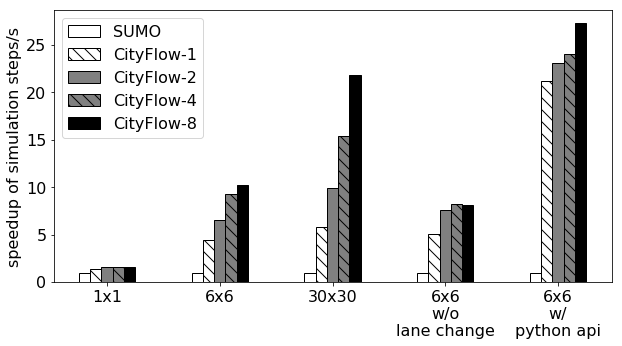}
  \caption{Speedup of \name compared to SUMO}
  \label{fig:performance}
\end{figure}
\subsection{Efficiency}

We compare the performance between SUMO and \name under different scenarios. The experiment runs on Intel(R) Xeon(R) CPU E5-2686 v4 @ 2.30GHz. As Figure \ref{fig:performance} shows, \name outperforms SUMO in all scenarios from small traffic to large traffic with single thread. The speedup is even more significant with more threads. We achieve about 25 times speedup on large scale $30\times 30$ road networks with tens of thousands of vehicles using 8 threads, which is 72 steps of simulation per second. Besides, \name shows better efficiency when retrieving information of the simulation via python interface. This is mainly because SUMO uses socket for interaction while \name uses \textit{pybind11} for seamless C++ and python integration.

\subsection{Effectiveness}
We evaluate the effectiveness of \name by comparing to SUMO because SUMO is already a widely-used traffic simulator and its effectiveness is acceptable by domain experts. We compare the average duration of vehicles (time for a vehicle to enter and leave the road network) under different traffic volume settings. As Table \ref{tab:diff} shows, the difference is within reasonable range.
\begin{table}[htbp]
  \centering
  \caption{Duration of vehicles under different traffic volume}
    \begin{tabular}{lrrrrr}
    Vehicles/Hour & 100   & 200   & 300   & 400   & 500 \\
    \midrule
    SUMO  & 40.76 & 41.57 & 42.75 & 44.08 & 45.93 \\
    \name & 40.79 & 41.58 & 42.62 & 43.84 & 45.45 \\
    Difference & 0.07\% & 0.04\% & 0.30\% & 0.54\% & 1.06\% \\
    \end{tabular}%
  \label{tab:diff}%
\end{table}%

\section{Demo Detail}
We plan to demonstrate \name in different traffic scenarios and show its capability to serve as reinforcement learning testbed. 

The demo consists of following parts:
\begin{itemize}
\item Simulating traffic in various scenarios, from synthetic grid scenarios to real world scenarios, and from small road networks with dozens of vehicles to large scale networks with tens of thousands of vehicles. 
\item Show the effectiveness the car-following model, intersection logic and lane change behavior of the simulator.
\item Show a complete reinforcement learning training episode of optimizing traffic signal plan. Participants can observe gradual improvement of traffic condition during the training.
\item Demo participants can control cycle length, green ratio of traffic signal and change the volume of traffic and see instant feedback of how the traffic condition would change. 
\end{itemize}

We have published a video on Youtube\footnote{https://youtu.be/qeE4hRmWONM}, which demonstrate the expected effect. The project is under active development and we are likely to add other features (e.g. more map options, vehicle controls) and demonstrate more functions at the conference.

No special hardware is required since we are demonstrating a software project (learning platform). We will bring our laptop. It would be great if a monitor is provided.

\section{Summary}
We propose \name, an efficient, multi-agent reinforcement learning environment for large scale city traffic scenario. Researchers can use it as a testbed for traffic signal control problem and conduct research on urban mobility. We will demonstrate the usage and some results of RL-controlled traffic signal plan. Also, we are actively developing the project and plan to support more RL scenarios like dynamic vehicle routing, policy of reversible lane or limited lane as well as open source the project in the near future.

\bibliographystyle{ACM-Reference-Format}
\balance
\bibliography{reference}

\end{document}